\documentclass{article}
\usepackage{spconf,amsmath,graphicx}
\usepackage{amsfonts}
\usepackage{multirow}
\usepackage{hyperref}
\usepackage{subfigure}
\usepackage{booktabs,setspace}
\usepackage{threeparttable}
\usepackage[subtle, tracking=normal]{savetrees}
\usepackage{marvosym}
\usepackage{setspace}

\usepackage{amssymb,CJK, indentfirst,balance,appendix}

\title{SLIDESPEECH: A LARGE SCALE SLIDE-ENRICHED AUDIO-VISUAL CORPUS}
\name{Haoxu Wang$^{1,2,3}$, Fan Yu$^{3}$, Xian Shi$^{3}$, Yuezhang Wang$^{3}$, Shiliang Zhang$^{3}$\textsuperscript{\Letter} \thanks{\Letter Corresponding author: sly.zsl@alibaba-inc.com.}, Ming Li$^{1,2}$}
\address{$^1$School of Computer Science, Wuhan University, Wuhan, China \\
$^{2}$Suzhou Municipal Key Laboratory of Multimodal Intelligent Systems,\\ Duke Kunshan University, Kunshan, China \\
    $^3$Speech Lab of DAMO Academy, Alibaba Group, China \\
}
\begin{document}
\ninept
\maketitle
\begin{abstract}
% The abstract should appear at the top of the left-hand column of text, about
% 0.5 inch (12 mm) below the title area and no more than 3.125 inches (80 mm) in
% length.  Leave a 0.5 inch (12 mm) space between the end of the abstract and the
% beginning of the main text.  The abstract should contain about 100 to 150
% words, and should be identical to the abstract text submitted electronically
% along with the paper cover sheet.  All manuscripts must be in English, printed
% in black ink.
Multi-Modal automatic speech recognition (ASR) techniques aim to leverage additional modalities to improve the performance of speech recognition systems. While existing approaches primarily focus on video or contextual information, the utilization of extra supplementary textual information has been overlooked. Recognizing the abundance of online conference videos with slides, which provide rich domain-specific information in the form of text and images, we release SlideSpeech, a large-scale audio-visual corpus enriched with slides. The corpus contains 1,705 videos, 1,000+ hours, with 473 hours of high-quality transcribed speech. Moreover, the corpus contains a significant amount of real-time synchronized slides.
In this work, we present the pipeline for constructing the corpus and propose baseline methods for utilizing text information in the visual slide context. Through the application of keyword extraction and contextual ASR methods in the benchmark system, we demonstrate the potential of improving speech recognition performance by incorporating textual information from supplementary video slides.
\end{abstract}

\section{Introduction}
\label{sec:intro}

 %This progress has not only led to the improvement and enhancement of ASR technology but has also facilitated the widespread adoption of intelligent devices. 

%to enhance recognition effectiveness in specific domains. The multi-modal information encompasses
% facial expressions\cite{avsv1,avsvhubert},

Recently, there has been a significant advancement in the development of automatic speech recognition (ASR) technology. Traditional methods based on Hidden Markov Models (HMM)\cite{gales2008application,povey2016purely} have been replaced by deep learning based techniques such as Connectionist Temporal Classification (CTC)\cite{graves2006connectionist,watanabe2017hybridctcatt}, Attention-based Encoder-Decoder (AED)\cite{attentionallyouneed,transfomer1,transformer2}, and Neural Transducer\cite{rnnt1,prunedrnnt}.

However, the development of a robust and generalized ASR model remains a challenge. Recognition performance tends to degrade in far-field or specific domain scenarios. To address this issue, scholars have explored the integration of multi-modal information such as lip movements\cite{deepavsr,avsrconformer,avsrhubert}, open domain information\cite{openavsr,seo2023avformer}, and contextual biasing list\cite{deepcontext,contextrnnt}, etc. This multi-modal information gives complementary lip movements, semantic images, and the context to improve the ASR results. %can improve the performance of far-field, By utilizing facial and lip movement data, speaker recognition (SV) performance can be improved, particularly in the presence of environmental noise. Lip movements are especially valuable for enhancing far-field speech recognition. In open domain scenarios, visual semantic information enhances semantic correspondence in recognition results. Utilizing contextual semantic information improves recognition outcomes in specific domains and dialog domain scenarios\cite{wei22c_interspeech}.

% In the past, there existed a certain amount of multimodal video datasets, part of which focused on face and lip movement domains, e.g., LRS2[], LRS3[], AV-Celeb[]; and part of them focused on open domain, e.g., HowTo100[], How2[], VisSpeech[];
% However, in addition to faces and semantic pictures, there is also multimodal text information in the video that is likely to be closely related to the speaker's current speech, and there are fewer articles that consider the use of text information in the video to go about improving the results of speech recognition. In online conference sharing scenarios, online education scenarios of free video, screen recording or live broadcasting, there are more slides synchronized with the speaker's speech, which contain a lot of rich textual domain information that has not been fully utilized. If only the asr technique is used, the recognition of proper nouns in the current domain will not be accurate enough. In the field of slides-assisted augmented speech recognition, some early articles used slides to build language models, some articles used complete static slides to extract rare words, and hot word models were used to improve speech recognition results;

In the past, there are many multi-modal video datasets, some of which focus on facial and lip movements, such as LRS2\cite{deepavsr}, LRS3\cite{afouras2018lrs3ted}, VoxCeleb2\cite{voxceleb2}, and CN-Celeb-AV\cite{cncelebav}. Others focus on open domains, such as HowTo100\cite{howto100M}, How2\cite{sanabria2018how2}, and VisSpeech\cite{openavsr}. However, in addition to facial expressions and semantic images, there is multi-modal textual information in the video that is closely related to the current speech of the speaker. Few articles consider utilizing textual information in videos to improve ASR results. In online conference sharing scenarios and online education scenarios where there are screen recordings or live broadcasts, there are often slides that are synchronized with the speaker's speech. These slides contain much rich textual information that has not been fully utilized. If only ASR technology is used, it may lead to the wrong recognition of proprietary entities in the current slide. In the field of ASR assisted by slides, some early papers use slides to build language models\cite{lecturelm,slideslm}, while others use complete static slides to extract rare words and improve results using a contextual bias ASR model\cite{sun22_interspeech}.

\begin{figure}[!t]
	\centering
        \vspace{-7pt}
	\includegraphics[scale=0.8]{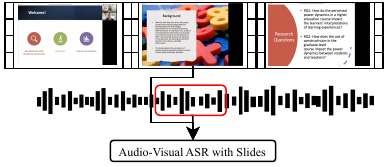}
        \vspace{-7pt}
	\caption{
		Example of our SlideSpeech using the Audio-Visual ASR Benchmark.
	}
	\label{slides}
        \vspace{-15pt}
\end{figure}

Previous works mainly focus on using static slides, with little consideration given to the high correlation between each short speech segment and the current slide. As shown in Fig. \ref{slides}, we propose to use real-time synchronized slide and speech stream for multi-modal ASR, as a "what you see is what you get" approach to improve the recognition of proprietary terms, and to avoid focusing solely on speech without paying attention to the textual information in the video slides.

%the real-time relationship between slide and speech, and without 

%The video and speech information in AMI is not completely synchronous
We investigate past database containing slides, and find only few database, such as AMI\cite{carletta2005ami} containing slide information. However, there are no time-synchronized recordings with slides sharing, and only few recordings containing slide images. 
To further study the improvement of ASR performance using real-time textual streams in multi-modal scenarios, we release SlideSpeech, a large-scale multi-modal audio-visual corpus with a significant amount of real-time synchronized slides. The key features of SlideSpeech include:

\begin{itemize}
    \item Abundance of slides in videos. These can be used for text-enhanced multi-modal ASR to correct the wrong recognized proprietary terms. % recognition performance of
    \item Sizeable. 1,705 videos with a total duration of 1,000+ hours, and 473 hours of transcribed speech with a confidence level above 95\%.
    \item Diverse. A diverse range of domain categories, with 22 classes. % It is characterized by its novelty and timeliness. distinct 
    \item Easy to expand or target specific scenarios. The slides in the videos also include relevant images and some facial information. This corpus can be applied for automatic subtitle generation in online education scenarios.
\end{itemize}
% Some of the recorded videos contain facial information.

We will present the pipeline used to construct the entire SlideSpeech corpus and discuss our approach to addressing the challenges of multi-modal ASR in the context of slides. Taking inspiration from the work of \cite{sun22_interspeech}, we develop a pipeline for text-based multi-modal ASR in synchronized slides and video scenarios. This pipeline incorporates various models, including text detection (TD), optical character recognition (OCR), semantic keyword extraction, and contextual bias ASR. Leveraging this pipeline, we establish a benchmark system based on SlideSpeech, enabling comparative analysis and further research.

\begin{figure}[!t]
	\centering
        \vspace{-7pt}
	\includegraphics[scale=0.65]{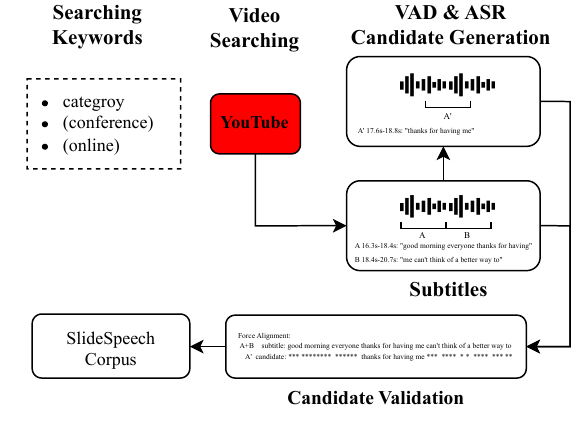}
        \vspace{-7pt}
	\caption{
		Diagram of our Creation pipeline.
	}
	\label{creationpipe}
        \vspace{-18pt}
\end{figure}

\vspace{-5pt}
\section{Creation Pipeline}
\label{sec:pipeline}
\vspace{-3pt}

As shown in Fig. \ref{creationpipe},  we introduce the detailed creation pipeline of our SlideSpeech corpus, including candidate video searching, candidate audio/text segments generation, and candidate validation.

\vspace{-10pt}
\subsection{Candidate Video Searching}
\vspace{-3pt}

% Our dataset primarily consists of video content sourced from the online video platform YouTube. Considering the prevalence of slide presentations in online conference scenarios, we design the retrieval keywords as specific categories + (conference) + (online). Our focus was on retrieving video playlists curated by uploaders or users, along with a small number of independent channel videos. We employ manual screening to collect slide videos, categorizing a video as a candidate if it featured slides for more than 50\% of its duration. Such videos are included in the download list. Additionally, playlists with 10 or more candidate videos or playlists comprising at least 80\% of candidates are also added to the download list. Ultimately, we utilized the yt-dlp\footnote{https://github.com/yt-dlp/yt-dlp} tool to download all the candidate videos. %, resulting in a total of xxx videos.

Our corpus primarily consists of video content from YouTube. We design retrieval keywords as specific categories + (conference) + (online), focusing on video playlists curated by uploaders or users, along with independent channel videos. Manual screening is conducted to collect slide videos, categorizing a video as a candidate if it features slides for over 50\% of its duration. Such videos are included in the download list. Playlists with 10 or more candidate videos or playlists comprising at least 80\% of candidates are also added to the download list. Finally, the yt-dlp\footnote{https://github.com/yt-dlp/yt-dlp} tool is used to download all the candidate videos.

\vspace{-7pt}
\subsection{Candidate Segments Generation}
\label{sec:candiate}
\vspace{-3pt}

Inspired by GigaSpeech\cite{gigaspeech} 
and WenetSpeech\cite{wenetspeech}, we use our in-house VAD and ASR systems to generate candidate transcripts for the downloaded videos. Our ASR system has shown exceptional performance on public benchmark platforms, consistently achieving accuracy rates exceeding 95\% across diverse testing scenarios and various public ASR databases. The audio from all videos is segmented using VAD, and the ASR system is then utilized to generate candidate transcripts for each segment. Thus, we obtain the audio/text segments.

% we utilized our internal commercial VAD and ASR systems to generate candidate transcripts for the downloaded videos. The transcription ASR system has demonstrated exceptional performance on the public benchmark platform, consistently achieving accuracy rates of over 95\% across diverse testing scenarios and many public ASR database. Our system employs VAD to segment long audio from all videos, followed by utilizing the ASR system to generate candidate transcript for each short audio segment. Here we get the audio/text segments.
\vspace{-7pt}
\subsection{Candidate Validation}
\vspace{-3pt}

While downloading the videos, we also obtain the automatically generated or user-uploaded subtitle files from YouTube. These files and the transcripts generated in Sec. \ref{sec:candiate} are used for candidate validation. However, due to their imprecise nature, we do not utilize the timestamps from the YouTube-generated subtitles. Nevertheless, the subtitle files generally cover all the text in the subtitles.
%We used these files along with the transcripts generated in the previous Sec \ref{sec:candiate} for candidates validation. However, as the timestamps in the YouTube-generated subtitles are intended for viewing purposes and are not precise, we did not utilize them. Nonetheless, the timestamps in the subtitle files generally covered all the text in the subtitles.

% Though the subtitles and ASR timestamps did not perfectly align, we still can use them for validation. For each short segment with candidate transcript generated by ASR system, we concatenate multiple subtitles based on their timestamps until they fully covered the current segment, thereby obtaining corresponding text subtitles and ASR candidate transcripts. As the concatenated subtitle time duration is longer than the segment duration, the subtitle text is longer than the candidate transcript. To align the longer subtitles and candidate transcripts, the Levenshtein algorithm\cite{} was applied. 

Though the subtitles and ASR timestamps do not perfectly align, we still can use them for validation. By concatenating multiple subtitles based on their timestamps, we ensure complete coverage of each short segment with a candidate transcript generated by the ASR system. Then, we obtain corresponding text subtitles and ASR candidate transcripts. As the concatenated subtitles' time duration exceeds the segment duration, the subtitle text is longer than the candidate transcript. To align the longer subtitles and candidate transcripts, we employ the Smith-Waterman algorithm\cite{pearson1991searching}. % Levenshtein algorithm\cite{}.

After aligning the two texts, we remove any unmatched words at the beginning and end of the subtitle text. We then computed the word error rate (WER) between the resulting subtitle text and the candidate transcript, and set $confidence = (1 - WER)$ for the current speech-text pair segment, which is used for subsequent database filtering.

\begin{figure*}[!t]
	\centering
        \vspace{-7pt}
	\includegraphics[scale=0.75]{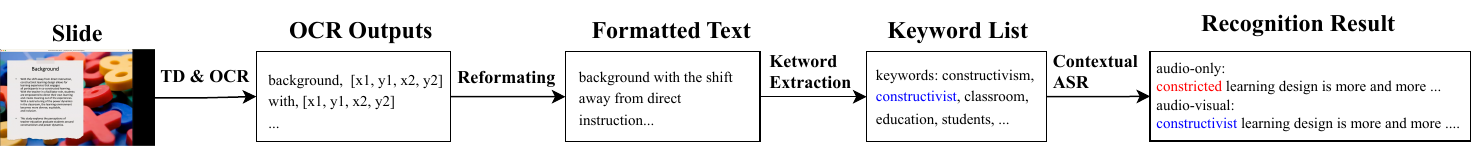}
        \vspace{-18pt}
	\caption{
		Diagram of our benchmark pipeline.
	}
	\label{benchmark}
        \vspace{-20pt}
\end{figure*}

\vspace{-5pt}
\section{The SlideSpeech Corpus}
\label{sec:slidespeech}
\vspace{-4pt}

In this section, we present the metadata, training sets, evaluation sets and the diversity of the corpus. For downloading the current corpus, please refer to the GitHub link provided\footnote{https://slidespeech.github.io/}. This corpus is not commercially available; it is only intended for academic research.

\vspace{-7pt}
\subsection{Metadata}
\vspace{-3pt}

% We provide all the metadata information of the videos, including original youtube channel, youtube playlist id, domain tags, and segments. We also provide the related timestamp, transcript, confidence for each segment. The original downloaded video files are provided and its format is 720p. The audio is also converted to 16k sample rate, single-channel, and 16-bit signed-integer format. In addition, we will provide preprocessed OCR results and the preprocessed extracted keywords for each segment.

We provide detailed metadata for the videos, including the original YouTube channel, YouTube playlist ID, domain tags, and segments. Each segment is accompanied by its corresponding timestamp, transcript, and confidence score. We also provide the scripts to download the videos. The downloaded video files are in 720p format, and the audios are converted to a 16k sample rate, single-channel, and 16-bit signed-integer format. We also offer preprocessed OCR results and extracted keywords for each segment.
\vspace{-7pt}
\subsection{Training Sets}
\vspace{-3pt}

The training set consists of 1659 videos totaling 1065.86 hours and is named as L. Moreover, a subset (S) is sampled from L, comprising 279 videos with a total duration of around 206 hours. All segments are automatically annotated with varying confidence. From S, another subset named S95 is created, containing segments with a confidence score above 95\%. S95 has an effective speech duration of about 161 hours. Similarly, a subset called L95 is formed from L, resulting in an effective speech duration of about 473 hours. Annotations with confidence levels below 95\% are retained in the original corpus for other academic purposes, such as self-supervised training.

\vspace{-7pt}
\subsection{Evaluation Sets}
\vspace{-3pt}

We provide development (dev) and test sets in addition to the training set. The dev set consists of 21 videos (5.07 hours), while the test set comprises 25 videos (8.75 hours). Unlike the training set, annotations in the dev and test sets are manually labeled. Additionally, after manually checking and random sampling 100 segments, we found that 94\% of segments in the dev and test sets contain slides. The dev and test sets are collected from YouTube, and we ensured no overlap with the training set regarding category and ID.
% covering 4.56 hours of speech, while the test set covers 7.96 hours.

% In addition to the training set, we have prepared validation and test sets. The validation set comprises a total of 21 videos with a duration of 5.07 hours, while the test set consists of 25 videos with a duration of 8.75 hours. In contrast to the training set, all annotations in the validation set are from manual labeling, covering a total speech duration of 4.56 hours, while the test set covers 7.96 hours of speech.

% Moreover, we have manually checked the validation and test sets, and random sample 100 segments and found that 94\% of the total segments in these sets contained slides. Our validation and test sets were also collected from the YouTube platform, and we verified that there is no overlap with the training set in terms of category and ID.

\begin{table}[!ht]
\vspace{-6pt}
\caption{The domian distribution with video counts and duration of the SlideSpeech.}
\centering
%\begin{threeparttable}[t]
% \vspace{3pt}

\setlength{\tabcolsep}{2pt}
\label{diversity}
\scalebox{0.9}{
\begin{tabular}{cccccc}
\hline
 Set &  Domian  & Num. & Dur.(h)  \\ \hline
\multirow{12}{*}{L}   & Computer Science  & 435      & 161.0     \\
   & Musical Instruments  & 23      & 29.1     \\
   & History  & 310      & 240.3      \\
   & Agriculture  & 67      & 69.3      \\
   & Animation  & 192      & 123.7      \\
   & Music  & 87      & 42.9      \\
   & Parenting  & 107      & 97.3      \\
   & Travel  & 108      & 93.6     \\
   & Life  & 43      & 36.3     \\
   & Talent  & 7      & 7.1     \\
   & English  & 20      & 17.0     \\
   & Other  & 260      & 148.4     \\
   \hline
\multirow{4}{*}{Dev} & Pet  & 5      & 1.57      \\
   & Health  &  6      & 1.36     \\
   & Dance  & 4      & 0.82     \\
   & Medical  & 6      & 1.32     \\
   \hline
\multirow{6}{*}{Test} & Fitness  & 3   & 1.98      \\
   & Design  &  5      & 1.09     \\
   & Traffic  & 3      & 0.84     \\
   & Education  & 5      & 1.88     \\
   & Tradition Culture  & 5      & 1.24     \\
   & Child  & 4      & 1.73     \\
 \hline
\end{tabular}
}
\vspace{-15pt}
\end{table}

\vspace{-7pt}
\subsection{Diversity}
\vspace{-3pt}
The categories in the retrieval keyword are utilized as labels for videos. Videos without a specific category or those manually identified as not belonging to any category were assigned to an "other" category. As shown in Table \ref{diversity}, the L set has 12 categories, the dev set has 4 categories, and the test set has 6 categories. Our corpus encompasses diverse categories, ensuring no bias towards a specific conference scenario. This makes it suitable for developing text-based multi-modal ASR techniques applicable to various scenarios.

% We used the keyword categories employed during retrieval as the category labels for the slide videos. For videos that lacked a specific category name or were identified manually as not belonging to a particular category, we assigned them to an "other" category. The L set includes 12 categories, the dev set includes 4 categories, and the test set includes 6 categories. Our slide video database features diverse categories, avoiding a bias towards a particular conference scenario. This makes it suitable for developing text-based multi-modal ASR techniques applicable to various scenarios.

% \vspace{-5pt}
\section{Benchmark System}
% \vspace{-3pt}

In this section, we present the text-based multi-modal ASR pipeline from the original video to the speech transcript. 

% \vspace{-10pt}
\subsection{Pipeline}
\vspace{-3pt}

As shown in Fig. \ref{benchmark}, for each speech segment, we extract the middle frame image and apply TD\cite{tdnet} and OCR\cite{ocrnet} models from the MMOCR toolkit\cite{mmocr2021} to extract the words in the slide. The words are then reformatted based on their coordinates to obtain formatted long text. Keyword extraction is performed using the KeyBert\cite{grootendorst2020keybert} technique, and contextual ASR is employed to enhance recognition for specific terms. Our benchmark system utilizes the contextual phrase prediction network (CPP network) ASR\cite{cpp}, as the contextual ASR model.

% For every speech segment, we extracted the middle frame image and used TD model\cite{} and OCR model\cite{} techniques to extract the words from the image, then reformate the words according to coordinates to get the formatted long text. We utilized KeyBert\cite{grootendorst2020keybert}\footnote{https://github.com/MaartenGr/KeyBERT} keyword extraction technique to extract keywords from the formatted text, and employed contextual ASR technique to improve the recognition performance for specific terms. We adopted the recent contextual phrase prediction network (CPP network) based ASR model\cite{cpp} as our contextual ASR model to build the benchmark system.
%

\vspace{-10pt}
\subsection{Contextual Bias ASR System}
\vspace{-3pt}

\begin{figure}[!t]
	\centering
        % \vspace{-7pt}
	\includegraphics[scale=0.87]{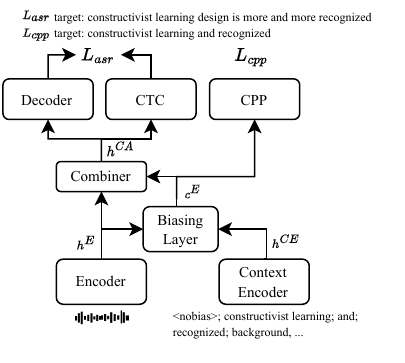}
        \vspace{-20pt}
	\caption{
		Diagram of Contextualized CTC/AED Model.
	}
	\label{model}
        \vspace{-17pt}
\end{figure}

We utilize Contextualized CTC/AED as our contextual ASR model, as depicted in Fig. \ref{model}. The model consists of a context encoder, a multi-head cross-attention based biasing layer, and a CPP network integrated with the traditional encoder. The contextual phrases are transformed into token sequences using byte-pair encoding (BPE). These sequences are then passed through a bidirectional LSTM, extracting the last time step's embedding ($h_{i}^{CE} \in R^{d}$) as the embedding for each contextual phrase $i$. Additionally, we include a special token \textless nobias\textgreater in the contextual biasing list before BPE encoding. This allows the model to handle cases where words from the bias list are absent in the speech, ensuring inference capability even without contextual phrases.

The biasing layer is a multi-head cross-attention mechanism that allows the speech embedding to capture contextual information from the contextual phrase embedding, modeling the relationship between speech and contextual phrases. By using the speech embedding ($h^{E}$) as the query and the contextual phrase embedding ($h^{CE}$) as the key and value, we obtain the contextual representation ($c^{E}$). The final contextual-enhanced speech embedding ($h^{CA}$) is obtained by fusing $c^{E}$ and $h^{E}$ using the combiner, defined as $h^{CA} = FeedForward([LayerNorm(h^{E}), LayerNorm(c^{E})])$. Additionally, the CPP network predicts contextual phrases present in the current speech. It consists of a linear projection layer and a shared CTC linear output layer with the main network. The CPP labels are generated by removing words from the original speech transcript that do not appear in the current contextual bias list, resulting in a CPP target text. During training, the context representation ($c^{E}$) is used to compute the CTC loss between $c^{E}$ and the CPP target text, serving as an auxiliary loss for the ASR model. This module is not required during testing.

% , including a followed by a feed-forward network (FFN)

%  By calculating the context representation $c^{E}$ using the following formula, and then  as per the following formula $h^{CA}$,

% \vspace{-10pt}
\vspace{-10pt}
\subsection{Training Bias List Generation}
\vspace{-3pt}

% In the original CPP paper\cite{cpp}, they use a simulation approach to randomly extract three phrases which containing 1-3 words from the original speech transcript, combined with some interfering words and \textless nobias\textgreater  to form the contextual bias list for training. Alternatively, we can extract a series of keywords from the real OCR text that are semantically related to the OCR text, forming a hotword list that maintains consistency between training and inference. These two methods can be used separately or in combination and can train an effective contextual ASR network.

In the original CPP paper\cite{cpp}, a simulation approach randomly selects three phrases containing 1-3 words from the speech transcript. These phrases are combined with distractors and \textless nobias\textgreater to form the contextual biasing list for training. Alternatively, we can extract semantically related keywords from the actual OCR text to form a biasing list. These methods can be used independently or combined to train an effective contextual ASR network.

% , ensuring consistency between training and inference.

\begin{table*}[!htb]\centering
% \begin{table*}[!h]

\footnotesize
    \caption{\label{resultstable} {\it Performance (\%) of the baseline model and the contextual asr benchmark. w/o K represents the case without Keyword, w K represents the case with Keyword. OCR, LR, Keyword columns denote the performance is calculated according to the OCR, LR or Keyword biasing list. U/B/R refers to the U-WER/B-WER/Recall metrics, respectively. %. The unit of recall rate is percentage (\%).  % refers to the all the recognized words in the slides as the bias list. R1 refers to the words whose recall rate below 40\% in the baseline recognition results but in the ocr text as the bias list. Keyword refers to the extracted keywords as the bias list.
    \vspace{+0.2em}
  }}
    % \begin{tabular}{l@{&}lllllllll}
\scalebox{0.85}{
    \begin{tabular}{cccccccccc}
    \toprule
    \multirow{2}*{\textbf{Model}} & \multirow{2}*{\textbf{Train} }& \multicolumn{4}{c}{\textbf{Dev}}  & \multicolumn{4}{c}{\textbf{Test}}  \\
    \cmidrule(lr){3-6} \cmidrule(lr){7-10}
    % \cmidrule(lr){2-7} \cmidrule(lr){8-10} \cmidrule(lr){11-13}  
     &  & WER & OCR(U/B/R) & LR(U/B/R) & Keyword(U/B/R) & WER & OCR(U/B/R) & LR(U/B/R) & Keyword(U/B/R) \\
    \midrule

    % 2D-ResNet34 \\
    % Baseline & S95 & 22.07 & 22.67/19.60/82.04 & 19.74/100/0.00 & 21.33/31.95/68.08 & 22.59 & 23.45/19.04/82.65 & 20.46/100/0.00 & 22.23/27.57/72.61 \\
    % Contextual w/o K & S95 & 22.01 & 22.43/20.29/81.38 & 20.17/83.85/16.30 & 21.19/33.05/67.04 & 22.42 & 23.23/19.09/82.48 & 20.81/80.83/19.89 & 21.98/28.50/71.66 \\
    % Contextual w K & S95 & 21.50 & 22.35/17.99/83.75 & 19.91/74.74/25.62 & 21.10/26.84/73.32 & 21.93 & 23.18/16.80/84.75 & 20.56/71.63/29.14 & 21.92/21.98/78.23 \\

    Baseline & S95 & 21.05 & 21.54/19.04/82.64 & 18.75/100/0.00 & 20.29/31.27/68.76 & 21.22 & 21.97/18.14/83.69 & 19.17/100/0.00 & 20.83/26.60/73.51 \\
    Contextual w/o K & S95 & 21.06 & 21.56/19.01/82.63 & 18.98/92.63/7.83 & 20.29/31.37/68.73 & 21.25 & 21.97/18.27/83.51 & 19.39/92.52/8.00 & 20.83/26.96/73.17 \\
    Contextual w K & S95 & 20.80 & 21.48/18.00/83.64 & 18.83/88.56/11.81 & 20.22/28.61/71.48 & 20.95 & 21.85/17.24/84.51 & 19.21/87.76/12.86 & 20.73/24.05/76.10 \\

    \midrule

    % Baseline & L95 & 14.03 & 14.76/11.02/90.04 & 12.66/100/0.00 & 13.86/16.31/83.69 & 14.18 & 15.07/10.54/90.35 & 12.97/100/0.00 & 14.20/13.87/86.22 \\
    % Contextual w/o K & L95 & 14.05 & 14.67/11.47/89.56 & 12.95/83.31/16.82 & 13.81/17.23/82.83 & 13.99 & 14.89/10.37/90.51 & 13.09/78.10/22.16 & 14.05/13.29/86.84 \\
    % Contextual w K & L95 & 13.65 & 14.60/\ 9.72/91.24 & 12.71/72.59/27.54 & 13.73/12.58/87.55 & 13.73 & 14.88/\ 9.045/91.85 & 12.98/67.01/33.42 & 14.04/\ 9.39/90.73 \\
    Baseline & L95 & 13.09 & 13.70/10.58/90.47 & 11.75/100/0.00 & 12.87/16.13/83.90 & 12.89 & 13.70/9.59/91.45 & 11.78/100/0.00 & 12.90/12.70/87.43 \\
    Contextual w/o K & L95 & 12.91 & 13.50/10.46/90.66 & 11.65/93.87/6.27 & 12.70/15.67/84.36 & 12.64 & 13.46/9.28/91.80 & 11.63/91.93/8.55 & 12.64/12.63/87.54 \\
    Contextual w K & L95 & 12.64 & 13.46/9.25/91.85 & 11.57/81.89/18.25 & 12.66/12.39/87.64 & 12.38 & 13.42/8.13/92.91 & 11.53/78.87/21.81 & 12.60/9.32/90.86 \\

    \bottomrule
\end{tabular}
}
% \vspace{-1.5em}
\end{table*}

\begin{figure*}[!t]
	\centering
        \vspace{-7pt}
	\includegraphics[scale=0.9]{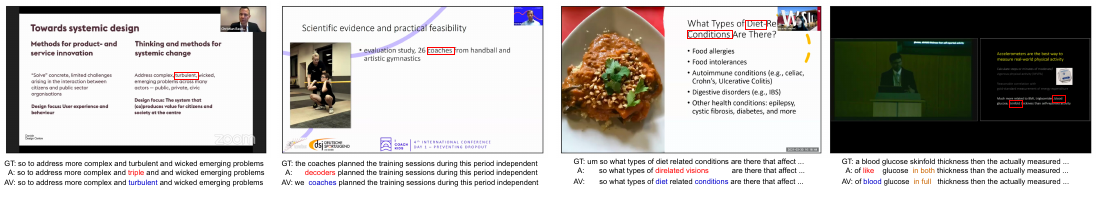}
        \vspace{-17pt}
	\caption{
		Qualitative results on the SlideSpeech dataset. We show the ground truth (GT), the recognition of the audio-only baseline (A) and benchmark system (AV). Note how the keywords of the slides help correct the recognition. Errors in the recognition compared to the GT are highlighted in red and the corrected words are highlighted in blue. The words are still wrong but in the slides are highlighted in orange. 
	}
	\label{qualitative}
        \vspace{-15pt}
\end{figure*}

\vspace{-5pt}
\section{Experiments and Discussion}
% \vspace{-2pt}

\vspace{-5pt}
\subsection{Experments Setup}
\vspace{-3pt}

% During the data preparation stage, we generate 80-dimensional FBank features using a 25ms window and a 10ms frame shift. Additionally, we apply SpecAugment\cite{specaug} with (F=30) and two time masks (T=40).
% Our baseline model is a conformer-based end-to-end model\cite{conformer}. The conformer model is an improved version of the transformer model, which not only retains the ability to model long-distance dependencies but also captures local information in speech. It has been widely used as a backbone for ASR systems. We utilize a set of 5k BPE tokens generated by the SentencePiece tokenizer, including an additional special token \textless nobias\textgreater. The model consists of 12 conformer blocks ($d^{ff} = 2048, H = 4, d^{att} = 256, CNN_{kernel} = 15$), and 6 transformer blocks in the decoder($d^{ff} = 2048, H = 4$). The objective function combines the CTC and attention objectives in a logarithmic linear combination, with a weight parameter $\lambda = 0.3$. Label smoothing is applied to the attention objective. Training of the conformer model was performed on four 16GB memory V100 RTX GPUs. The maximum trainable epoch was set to 70, with a mini-batch size of 22 million acoustic feature bins. We used the Adam optimizer without weight decay and applied the Noam learning rate scheduler with 15k warmup steps and a learning rate of 0.002. The final model was obtained by averaging the last 10 best checkpoints.

During data preparation, we generate 80-dimensional FBank features with a 25ms window and a 10ms frame shift. SpecAugment\cite{specaug} is applied with frequency and time masks (F=30, T=40).
Our baseline model is a conformer-based end-to-end model\cite{conformer}, an improved version of the transformer model that captures both long-distance dependencies and local information in speech. It consists of 12 conformer encoder blocks ($d^{ff} = 2048, H = 4, d^{att} = 256, CNN_{kernel} = 15$) and 6 transformer decoder blocks ($d^{ff} = 2048, H = 4$). We utilize a set of 5k BPE tokens generated by the SentencePiece tokenizer and an added special token \textless nobias\textgreater. The objective function combines CTC and attention objectives with a logarithmic linear combination using a weight parameter $\lambda = 0.3$. Label smoothing is applied to the attention objective. Training is performed on four 16GB memory V100 RTX GPUs, with a maximum trainable epoch of 70 and a mini-batch size of 22 million acoustic feature bins. We use the Adam optimizer and apply the Noam learning rate scheduler with 15k warmup steps and a learning rate of 0.002. The final model is obtained by averaging the top 10 best checkpoints.

% without weight decay 
% a three-fold speed perturbation and
%  with an output dimension of 256 and a kernel size of 31 in the encoder
% Both the encoder and decoder have 4 attention heads with a feed-forward unit dimension of 2048.

% In the contextualized version, the context encoder consists of a 1-layer bidirectional LSTM (BLSTM) with a dimensionality of 256 and a linear layer. The biasing layer is a multi-head attention (MHA) layer with an embedding size of 256 and 4 attention heads. The first linear layer in the context prediction network maintains a 256-dimensional input and output and follow a tanh activation, while the second linear layer projects the input onto the vocabulary size and shares parameters with the CTC linear layer. We first train the contextual model using the simulated biasing list and then finetune on the combination of simulated and keywords biasing list with a mix ratio of 0.5. All experiments were implemented using the ESPnet end-to-end speech processing toolkit\cite{watanabe2018espnet}.

In the contextualized version, the context encoder consists of a 1-layer bidirectional LSTM (BLSTM) with a dimensionality of 256 and a linear layer. The biasing layer is a multi-head attention (MHA) layer with an embedding size of 256 and 4 attention heads. The context prediction network consists of two linear layers. The first linear layer has a 256-dimensional input and output with a tanh activation, while the second linear layer projects the input onto the vocabulary size and shares parameters with the CTC linear layer. We initially train the contextual model using the simulated biasing list and then finetune it on a combination of simulated and keyword biasing lists with a 0.5 mix ratio. We initially only train the contextual part and freeze the others from the pretrained baseline model and unfreeze all when finetune it on the mixture of the simulated and keyword biasing lists. The experiments were conducted using the ESPnet end-to-end speech processing toolkit\cite{watanabe2018espnet}.
% For all training parts, 0.9 of the samples and 0.9 of the batches have only \textless nobias\textgreater as the biasing list.

\vspace{-7pt}
\subsection{Results}
\vspace{-3pt}

% We evalute the results based on WER, biased word error rate (B-WER),  unbiased word error rate (U-WER) and the recall of the words which are both in biasing list and the transcript. The U-WER is calculated specifically on words that are not included in the biasing list, while the B-WER is computed on words that are included in the biasing list. In cases of insertion errors, if the inserted phrase belongs to the biasing list, it contributes to the B-WER calculation; otherwise, it contributes to the U-WER calculation. 

We evaluate the results using WER, biased word error rate (B-WER), unbiased word error rate (U-WER), and the recall of words in both the biasing list and transcript. U-WER is calculated explicitly for words not in the biasing list, while B-WER is computed for words in the biasing list. In cases of insertion errors, if the inserted phrase is from the biasing list, it contributes to the B-WER calculation; otherwise, it contributes to the U-WER calculation.

%The biasing list used during inference and the reference biasing list used during calculating metrics can be different. We consistently used a biasing list of 50 keywords during inference. We reported the results for different reference biasing lists in the table \ref{resultstable}. The OCR biasing list refers to using all the words in the current slide as the biasing list, representing the upper limit of word assistance information that can be provided by the current slide. The R1 biasing list refers to the subset of words that are present in the slide but are not recognized correctly by the baseline audio-only model, representing the upper limit of utilizing slide information for correction. The Keyword biasing list refers to the list of 50 keywords used during inference, which is closely related to the semantic information of the slide. The results of Keyword biasing list represent the performance of the contextual ASR model. % Successfully recognizing these words can significantly improve the user experience. 

The biasing list used during inference and the reference biasing list used during calculating metrics can be different. We use a biasing list of 50 keywords during inference. Table \ref{resultstable} presents the results for various reference biasing lists. The OCR biasing list includes all the words on the slide, representing the upper limit of word assistance information that the current slide can provide. The Low-Recall (LR) biasing list comprises words present on the slide but not correctly recognized by the baseline audio-only model, representing the upper limit of utilizing slide information for correction. Lastly, the Keyword biasing list consists of the 50 keywords used during inference, closely related to the semantic information of the slide. The results of the Keyword biasing list indicate the contextual ASR model's performance.

Our baseline audio-only model trained on S95/L95 achieves 21.05/13.09 and 21.22/12.89 WER on the Dev and Test Set, respectively. On the OCR biasing list, it achieves 19.04/10.58 B-WER with a recall of 82.64\%/90.47\%, and 18.14/9.59 B-WER with a recall of 83.69\%/91.45\%. As mentioned earlier, the LR metrics represent the potential to correct recognition errors using slide information. The Keywords biasing list contains the words that carry semantic information related to the slide. The baseline trained on S95/L95 achieved 31.27/16.13 with a recall of 68.76\%/83.90\% on the Dev set, and 26.60/12.70 B-WER with a recall of 73.51\%/87.43\% on the Test set.
%The performance of the baseline model degrades when using the R1 biasing list, which represents the potential to correct recognition errors using slide information. 
% The baseline trained on L95 achieves 14.03 and 14.18 WER. 

% However, there were still some words that were not successfully recognized but could be referenced in the slide content, resulting in an R1 biasing list.
% , incorporating the text-based audio-visual ASR method
% On the Keyword metric,
Compared to the baseline, the contextual ASR without the Keyword biasing list shows comparable results in WER. Using the 50 Keyword biasing list during inference, the WER of contextual ASR trained on S95/L95 improves to 20.80/12.64 and 20.95/12.38 on the Dev and Test sets. Moreover, improvements are observed in OCR, LR, and Keyword cases. In the case of Keyword, the B-WER of the model trained on S95/L95 improves to 28.61/12.39 and 24.05/9.32 on the Dev and Test sets, with a recall improvement to 71.48\%/87.64\% and 76.10\%/90.86\%. These results demonstrate the potential for enhancing specialized terms using additional slide information on this corpus. It should be noted that our main contributions include providing an open-source corpus and offering baseline methods for this corpus. The OCR metric is used to compare the performance of using slide information, and the Keyword metric is used to compare the performance of the contextual ASR model.
% , consistent with the original paper
%  (a relative improvement of xx and xxx)

% \subsection{Qualitative Results}

\noindent \textbf{Qualitative Results}: 
Further improvement results are presented in Fig. \ref{qualitative}. Leveraging textual information from slides can effectively enhance the recognition performance of ASR systems, especially for recognizing specialized and proper nouns. The dataset used in this study and the provided baseline system serves as an essential research reference for text-based audio-visual ASR.
% Further improvement results are presented in Fig. \ref{qualitative}. It is evident that leveraging textual information from slides can effectively enhance the recognition performance of ASR systems, especially for recognizing specialized and proper nouns. The dataset used in this study, along with the provided baseline system, serves as an important research reference for text-based audio-visual ASR.

\vspace{-5pt}
\section{Conclusion}
\vspace{-4pt}

In this work, we release SlideSpeech, which is a large-scale audio-visual corpus enriched with slides. The corpus contains a significant amount of real-time synchronized slides. We introduce the creation pipeline and the details of the corpus. We also present a benchmark system for this slide-enriched corpus. The experiment results demonstrate the potential for enhancing specialized terms using additional slide information on this corpus.

\footnotesize
\bibliographystyle{IEEEbib}
\bibliography{strings,refs}
% \end{spacing}

\end{document}